\documentclass[nofootinbib,tightenlines]{revtex4-2}
\usepackage{amssymb,amsmath,amsthm,graphicx,subfigure,float,cancel,braket,comment,tipa,bm,dsfont,graphicx,hyperref}
\usepackage{xcolor}
\usepackage{mathrsfs,enumitem}
\hypersetup{
	colorlinks,
	linkcolor={red!50!black},
	citecolor={blue!50!black},
	urlcolor={blue!80!black}
}
\usepackage{mathptmx}
\usepackage[T1]{fontenc}

\begin{document}
	\title{Isometries of N=1 4D supergravity}
	\author{Nephtalí Eliceo Martínez Pérez}
	\email{nephtali.martinezper@alumno.buap.mx}
	\author{Cupatitzio Ramírez Romero}
	\email{cramirez@fcfm.buap.mx}
	\address{Facultad de Ciencias Físico Matemáticas, Benemérita Universidad Autónoma de Puebla}
	
\begin{abstract}
Continuous symmetries of spacetime such as spatial homogeneity and isotropy are rigorously defined using the concept of isometries and Killing vectors. In supergravity, the metric, or rather the tetrad, is not a standalone entity, but is part of a multiplet containing also the Rarita-Schwinger spinor-vector. Thus, we pursue a generalization of the Killing equations that is in harmony with the tenets of supergravity. Using a superfield approach, we derive one such generalization of the Killing equations encompassing the whole supergravity multiplet. A relaxation of the spinor-vector equations is required to allow for a non-vanishing isotropic Rarita-Schwinger field.
\end{abstract}

\maketitle
\section{Introduction}
N=1 supergravity in four dimensions is an extension of general relativity in which the gravitational field and a Rarita-Schwinger spinor-vector form a closed set under gauged supersymmetry transformations \cite{wessbagger,freedman}. As with any theory of fermions, supergravity is ultimately designed as a quantum theory. For example, the quantum field theory of linearized supergravity describes spin-2 and spin-$\frac{3}{2}$ particles on flat spacetime \cite{west}.

Another form of quantum supergravity finds application in quantum cosmology \cite{deathbook,moniz,moniz2,macias}. In this branch, it is common the introduction of symmetries at the classical level to obtain more tractable models, e.g., spatially homogeneous and isotropic, or FRW, models \cite{hartle,halliwell}. This method is readily implemented in scenarios of modified gravity such as $f(R)$-theories by evaluating the action at the FRW metric \cite{hawkingluttrell,vilenkin85}. With supersymmetry, however, things are not that straightforward and there are different approaches to supersymmetric cosmology. For example, a 1D action with N=4 local supersymmetry can be derived by evaluating the supergravity action for a k=1 FRW tetrad and a homogeneous Rarita-Schwinger field \cite{death88,death92,asano}. On the other hand, k=0 and k=1 FRW models with N=2 local supersymmetry have been obtained with superfield generalizations of the one-dimensional FRW actions, where the scale factor is treated as a matter field \cite{obregon1,obregon2,garcia,universe8080414,higherderivatives}. While these approaches successfully provide specific models, a general formulation of symmetries has not been given.

In Einstein gravity, continuous symmetries of the spacetime manifold, called isometries, are determined with the Killing equations \cite{torres}. In supergravity, the tetrad, containing the spacetime geometry and the local frame of reference, transforms into the Rarita-Schwinger field and vice versa. This suggests an extension of the Killing equations telling how a symmetry of the tetrad transfers to the other components of the supermultiplet. In this work, we explore an extension of the Killing equations, attending the basic requirement that one recovers the standard bosonic expressions in the limit of vanishing fermions. We emphasize that we are interested in defining isometries of supergravity regardless of the field equations. For example, in cosmology, the generic FRW metric admits six spacelike Killing vectors \cite{ellis}, but it is only a solution of general relativity when the scale factor satisfies the Friedmann equations. Thus, in contrast to the usual classical solutions of supergravity \cite{freedman}, we are considering classical configurations with non-vanishing fermions \cite{finkelstein}.

This paper is organized as follows. In Section \ref{sec2}, we review the Killing equations in the tetrad formulation (\ref{killingtetrad}). In Section \ref{sec3} the tetrad Killing equations are promoted to the level of superspace in terms of the vielbein (\ref{superkilling}) and tested with flat superspace. Next, to deal directly with the components of supergravity, we evaluate the superfield equations in the Wess-Zumino gauge. In Section \ref{sec4}, we describe the first prospect of the supergravity Killing equations (\ref{sugrakilling}). Since the spinor-vector equations (\ref{spinorvector}) yield, in particular, a vanishing isotropic Rarita-Schwinger field, we consider alternative equations (\ref{spinormetric}), while preserving supergravity. Finally, in Section \ref{sec6}, we draw conclusions and comment on open questions. Some useful results and complementary computations are collected in Appendix \ref{app1}.

\section{Isometries of spacetime}\label{sec2}
For definiteness, spacetime is a four dimensional pseudo-Riemannian differentiable manifold. The geometry of spacetime can be recovered from the auxiliary tetrad fields $e_m^{\ a}(x)$, which are related to the metric by \cite{bojowald},
\begin{align}\label{metrictetrad}
	g_{mn}(x)=\eta_{ab} e_m^{\ a}(x) e_n^{\ b}(x),
\end{align}
with the Minkowski metric $\eta_{ab}=\text{diag}(-1,1,1,1)$. 

The tetrad fields undergo spacetime coordinate transformations and local Lorentz transformations, respectively,
\begin{align}\label{two}
e_{m'}^{\ a}(x')=\frac{\partial x^n}{\partial x'^m} e_n^{\ a}(x), &&
e_m^{\ a'}(x)=e_m^{\ b}(x) \Lambda_b^{\ a}(x).
\end{align}
For a given $g_{mn}(x)$, there is an entire equivalence class of tetrads, each related to one another by a local Lorentz transformation, that satisfy (\ref{metrictetrad}).

Isometries comprise coordinate transformations $x^m\to x'^m(x)$ that leave the form of the metric invariant \cite{inverno},
\begin{align}\label{first}
	g'_{mn}(x)=g_{mn}(x),
\end{align}
where the left-hand side is the transformed tensor evaluated at argument $x$. Expressing (\ref{first}) in terms of tetrads we learn that there is $\Lambda_b^{\ a}(x)$ such that, in the notation of (\ref{two}),
\begin{align}\label{class}
	e_{m'}^{\ a}(x)=e_m^{\ a'}(x).
\end{align}
That is, the transformed tetrad belongs to the class of the original tetrad.

For small transformations modulated by the real parameter $h$ taking values in a neighborhood of zero,
\begin{align}
x'^m=x^m+h \xi^m(x), && \Lambda_b^{\ a}(x)=\delta_a^{\ b}+h L_b^{\ a}(x),
\end{align}
taking the limit of (\ref{class}) as $h\to 0$, we obtain
\begin{align}\label{killingtetrad}
\delta_\xi e_m^{\ a}\equiv -\xi^l(x) \partial_l e_m^{\ a}(x)-e_l^{\ a}(x) \partial_m \xi^l(x)=e_m^{\ b}(x) L_b^{\ a}(x),
\end{align}
which can be taken as the Killing equations in tetrad form \cite{chinea}. Indeed, multiplying (\ref{killingtetrad}) by $e_{na}$ and adding to it the expression with $m$, $n$ interchanged, yields the usual Killing equations in terms of the metric \cite{torres},
\begin{align}\label{killingmetric}
\xi^l \partial_l (e_m^{\ a} e_{na})+e_l^{\ a} e_{na} \partial_m \xi^l+e_m^{\ a} e_{la} \partial_n \xi^l=0.
\end{align}

Performing a local Lorentz transformation on (\ref{killingtetrad}), yields
\begin{align}
\delta_\xi (e_m^{\ b} L_b^{\ a})=e_m^{\ c} \Lambda_c^{\ b} (\Lambda^{-1} L\Lambda-\Lambda^{-1} \xi^l \partial_l \Lambda)_b^{\ a} \equiv e_m^{\ c} \Lambda_c^{\ b} K_b^{\ a}.
\end{align}
It can be shown that $K_{ba}=-K_{ab}$, therefore, the $\xi^l$ are also Killing vectors of the transformed tetrad, as expected from (\ref{killingmetric}).

Since the tetrad is not symmetric, there are six more equations in (\ref{killingtetrad}) than in (\ref{killingmetric}). They determine the Lorentz generators, which depend on both $\xi^m$ and the chosen tetrad, 
\begin{align}\label{solvelorentz}
	L_b^{\ a}(x)=-e_b^{\ m} (\xi^l \partial_l e_m^{\ a}+e_l^{\ a} \partial_m \xi^l)
\end{align}

\section{Isometries of superspace}\label{sec3}
Superspace is an extension of spacetime in which supersymmetry transformations arise as coordinate transformations of the additional anticommuting dimensions. A generic system of local coordinates on superspace is $z^M=(z^m, z^\mu, z_{\dot{\mu}})=(x^m,\theta^\mu,\bar{\theta}_{\dot{\mu}})$; $x^m$ are ordinary spacetime coordinates, whereas $\theta^\mu, \bar{\theta}^{\dot{\mu}}=(\theta^\mu)^*$ are anticommuting Grassmann variables with two-component spinor indices \cite{wessbagger}. A summary of conventions and other important results following \cite{wessbagger} are given in Appendix \ref{appa}.

The superspace analogue of the tetrad fields are the vielbein superfields, $E_M^{\ A}(z)$, with local indices $A=(a,\alpha, \dot{\alpha})$ transforming in the four-vector and Weyl representations of the Lorentz group. The basic transformations of the vielbein are general  coordinate transformations (GCT's),
\begin{align}\label{transsuper}
	E_{M'}^{\ \ A}(z')&=(\partial'_M z^N) E_N^{\ A}(z)\equiv (\partial'_M x^n) E_n^{\ A}(z)+(\partial'_M \theta^\nu) E_\nu^{\ A}(z)+(\partial'_M \bar{\theta}_{\dot{\nu}}) E^{\dot{\nu} A}(z),
\end{align}
with partial derivatives $\partial_m=\partial/\partial x^m$, $\partial_\mu=\partial/\partial \theta^\mu$, $\partial^{\dot{\mu}}=\partial/\partial \bar{\theta}_{\dot{\mu}}$, and local Lorentz transformations (LLT's),
\begin{align}
	E_M^{\ \ A'}(z)&=E_M^{\ B}(z) \Lambda_B^{\ A}(z)\equiv E_M^{\ b}(z) \Lambda_b^{\ A}(z)+E_M^{\ \beta}(z) \Lambda_\beta^{\ A}(z)+E_{M\dot{\beta}}(z) \Lambda^{\dot{\beta} A}(z),
\end{align}
with $\Lambda_B^{\ A}(z)$ non-vanishing only for the irreducible representations: $\eta_{cd}=\Lambda_c^{\ a}(z) \eta_{ab} \Lambda_d^{\ b}(z)$, $\epsilon_{\gamma \delta}=\Lambda_\gamma^{\ \eta}(z) \epsilon_{\eta \kappa} \Lambda_\delta^{\ \kappa}(z)$, etc.

Now, since superfield expressions yield most times to the proper supersymmetric generalizations of bosonic theories, we begin our search of the supergravity Killing equations by promoting (\ref{killingtetrad}) to superfield equations as follows,
\begin{align}\label{superkilling}
\delta_\xi E_M^{\ A}(z) \equiv -\xi^L(z) \partial_L E_M^{\ A}(z)-(\partial_M \xi^L(z)) E_L^{\ A}(z)=E_M^{\ B}(z) L_B^{\ A}(z),
\end{align}
where $\xi^M(z)$ and $L_B^{\ A}(z)$ are the generators the infinitesimal transformations,
\begin{align}\label{infsuper}
z'^M=z^M+h \xi^M(z), && \Lambda_B^{\ A}(z)=\delta_B^{\ A}+h L_B^{\ A}(z).
\end{align}
Thus, we are defining isometries of superspace as coordinate transformations whose effect on the vielbein amounts to a local Lorentz transformation.

Let's test the suitability of (\ref{superkilling}) with flat superspace given by \cite{wessbagger}
\begin{align}\label{flat}
E_M^{\ A}(z)=\begin{bmatrix}
\delta_m^{\ a} & 0 & 0 \\
-i \sigma^a_{\mu \dot{\mu}} \bar{\theta}^{\dot{\mu}} & \delta_\mu^{\ \alpha} & 0 \\
i \epsilon^{\dot{\mu} \dot{\nu}} \theta^\nu \sigma^a_{\nu \dot{\nu}} & 0 & \delta^{\dot{\mu}}_{\ \dot{\alpha}}
\end{bmatrix}
\end{align}
The associated 4-geometry is Minkowski spacetime; the spinor-vector and the connection vanish. 

Substituting (\ref{flat}) into (\ref{superkilling}) and simplifying we obtain the following set of equations. For $(M,A)=((m, \mu),a)$,
\begin{subequations}
	\begin{align}
	-(\partial_m \xi^n(z)) \delta_n^{\ a}&=\delta_m^{\ b} L_b^{\ a}(z), \label{normal} \\
	-\bar{\xi}_{\dot{\nu}}(z) \partial^{\dot{\nu}} E_\mu^{\ a}(\bar{\theta})-(\partial_\mu \xi^n(z)) \delta_n^{\ a}-(\partial_\mu \xi^\nu(z)) E_\nu^{\ a}(\bar{\theta})&=E_\mu^{\ b}(\bar{\theta}) L_b^{\ a}(z). \label{rec1}
	\end{align}
\end{subequations}
Equations $(\dot{\mu},a)$ are the complex conjugate of (\ref{rec1}). For $(M,A)=((m,\mu,\dot{\mu}),\alpha)$, we obtain
\begin{align}
-(\partial_m \xi^\nu(z)) \delta_\nu^{\ \alpha}=0, && -(\partial_\mu \xi^\nu(z)) \delta_\nu^{\ \alpha}=\delta_\mu^{\ \gamma} L_\gamma^{\ \alpha}, && -(\partial^{\dot{\mu}} \xi^\nu(z)) \delta_\nu^{\ \alpha}=0. \label{fg2}
\end{align}

Assuming $\xi^n$ and $L_b^{\ a}$ are functions of spacetime only, the solution of (\ref{normal}) can be written as linear combination of generators of Poincaré transformations: spacetime translations $\partial_m$, spatial rotations $J_i=\epsilon_{ij}^{\ \ k} x^j \partial_k$ and boosts $K_i=x^i \partial_0-x^0 \partial_i$, for which $L_b^{\ a}(\partial_m)=0$, $L_b^{\ a}(J_i)=-\epsilon_{ij}^{\ \ k} \delta_b^{\ j} \delta_k^{\ a}$ and $L_b^{\ a}(K_k)=\delta_b^{\ 0} \delta_k^{\ a}-\delta_b^{\ k} \delta_0^{\ a}$, respectively. Next, equations (\ref{fg2}) hold if
\begin{align}\label{integrating}
\xi^\nu(z)=-\theta^\rho \delta_\rho^{\ \beta} L_{\beta}^{\ \alpha}(\bm \xi) \delta_\alpha^{\ \nu}, && \bar{\xi}_{\dot{\nu}}(z)=-\bar{\theta}_{\dot{\rho}} \delta^{\dot \rho}_{\ \dot{\beta}} L^{\dot{\beta}}_{\ \dot{\alpha}}(\xi) \delta^{\dot \alpha}_{\ \dot{\nu}}
\end{align}
with the spinor Lorentz generators given in (\ref{generators}). Finally, solving for $\partial_\mu \xi^l$ from (\ref{rec1}), using (\ref{integrating}) and the spinor Lorentz generators (\ref{generators}), we obtain $\partial_\mu \xi^l=0$, in agreement with the assumption that $\xi^n(z)=\xi^n(x)$.

The fact that flat superspace admits ten independent Killing vectors, but has a vanishing Rarita-Schwinger field, can be posed as maximum symmetry spoiling supergravity. The question then is whether or not less symmetry allows a simpler but nontrivial version of supergravity.

On the other hand, (\ref{superkilling}) have as input the full $\theta$-expansion of the vielbein, which is generally not available\footnote{The full series expansion can  been given in the covariant W-Z gauge using the so-called new $\Theta$-variables \cite{ramirez}}. Thus, in the next section, we evaluate equations (\ref{superkilling}) in a particular gauge, thereby obtaining equations for the supergravity multiplet only.

\subsection{The supergravity multiplet}
Infinitesimal transformations such as (\ref{superkilling}) can be implemented with parameters carrying either superspace or local indices, which are related by the vielbein,
\begin{align}\label{local}
\xi^A(z)=\xi^M(z) E_M^{\ A}(z), && \xi^M(z)=\xi^A(z) E_A^{\ M}(z).
\end{align}

In particular, the left-hand side of (\ref{superkilling}) can be written in local form as follows \cite{wessbagger},
\begin{align}\label{translaw}
\delta_\xi E_M^{\ A}&=-\xi^L \partial_L E_M^{\ A}-\partial_M (\xi^L E_L^{\ A})+(-)^{lm} \xi^L \partial_M E_L^{\ A} \nonumber \\
&=-\xi^L \big(\partial_L E_M^{\ A}-(-)^{lm} \partial_M E_L^{\ A}-(-)^{bm} E_L^{\ B} \phi_{MB}^{\ \ \ A}\big)-(\partial_M \xi^A+(-)^{bm} \xi^L E_L^{\ B} \phi_{MB}^{\ \ \ A}) \nonumber \\
&=-\xi^L T_{LM}^{\ \ \ A}+E_M^{\ B} \xi^L \phi_{LB}^{\ \ \ A}-\mathscr{D}_M \xi^A,
\end{align}
using the torsion components (\ref{torsion}). A Lorentz covariant expression can be achieved by performing a field-dependent Lorentz transformation, $L_B^{\ A}=-\xi^L \phi_{LB}^{\ \ A}$, in addition to $\xi^A$ in (\ref{translaw}). These are called supergauge transformations and they are used to generate supergravity transformations \cite{wessbagger}.

Superfields introduce more component fields than necessary for a consistent theory. One gets rid of them with covariant constraints such as (\ref{const}) on the torsion components, or by fixing the gauge. For example, higher $\theta$-components of the infinitesimal generators (\ref{translaw}) can be used to fix the lowest components of the vielbein. In particular, 
\begin{subequations}\label{wzgauge}
	\begin{align}
	E_M^{\ \ A}(z)|&=\begin{bmatrix}
	e_m^{\ a}(x)  & \frac{1}{2} \psi_m^{\ \alpha}(x) & \frac{1}{2} \bar{\psi}_{m \dot{\alpha}}(x) \\
	0 & \delta_\mu^{\ \alpha} & 0\\
	0 & 0 & \delta^{\dot{\mu}}_{\ \dot \alpha}
	\end{bmatrix}, \label{wzviel} \\
	\phi_{MA}^{\ \ \ B}(z)|&
	=\begin{bmatrix}
	\omega_{mA}^{\ \ \ B}(x) & 0 & 0
	\end{bmatrix}, \label{connection}
	\end{align}
\end{subequations}
where $|$ denotes evaluation at $\theta=0=\bar{\theta}$, is known as the Wess-Zumino (W-Z) gauge.

The N=1 minimal supergravity multiplet contains the tetrad $e_m^{\ a}(x)$, the Rarita-Schwinger spinor-vector $\psi_m^{\ \alpha}(x)$, its complex conjugate $\bar{\psi}_m^{\ \dot{\alpha}}(x)$, and six real auxiliary fields: A complex scalar $M(x)$ and a real four-vector $b_a(x)$. The transformation rules under local supersymmetry transformations parameterized by $\zeta^\alpha(x)$ are
\begin{subequations}\label{sugratrans}
	\begin{align}
	\delta_\zeta e_m^{\ a}&=i (\psi_m \sigma^a \bar{\zeta}-\zeta \sigma^a \bar{\psi}_m), \label{sugratetrad}\\
	\delta_\zeta \psi_m^{\ \alpha}&=-2 \mathscr{D}_m \zeta^\alpha+\tfrac{i}{3} e_m^{\ c} (3 b_c \zeta^\alpha+b_d (\zeta \sigma^d \bar{\sigma}_c)^\alpha-M (\bar{\zeta} \bar{\sigma}_c)^\alpha), \label{sugraspinor} \\
	\delta_\zeta M&=-2 \zeta \sigma^{ab} \psi_{ab}+i \zeta \sigma^a \bar{\psi}_a M-i b^a \zeta \psi_a, \label{sugram} \\
	\delta_\zeta b_a&=3 (\zeta^\alpha \mathscr{D}_\alpha G_a(z)|+\bar{\zeta}_{\dot{\alpha}} \bar{\mathscr{D}}^{\dot{\alpha}} G_a(z)|), \label{sugravector}
	\end{align}
\end{subequations}
Expression (\ref{sugravector}) can be further developed, but this form is sufficient for our needs. For more details about the superspace formulation of N=1 supergravity, we refer the reader to reference \cite{wessbagger}.

\subsection{Local spacetime diffeomorphism}
Using (\ref{translaw}), the superfield equations (\ref{superkilling}) become
\begin{align}\label{newkilling}
-\mathscr{D}_M \xi^A-\xi^B T_{BM}^{\ \ \ A}+E_M^{\ B} \xi^C \phi_{CB}^{\ \ \ A}=E_M^{\ B} L_B^{\ A}.
\end{align}
We proceed to evaluate this expression in the W-Z gauge (\ref{wzgauge}) for the transformation parameters
\begin{align}\label{prev}
\xi^A|=(\xi^a(x), \xi^\alpha(x), \bar{\xi}^{\alpha}(x))=(\xi^a(x),0,0).
\end{align}

As shown in Appendix \ref{appb}, higher-order $\theta$-components of $\xi^A$ and $L_B^{\ A}$ can be used to preserve the W-Z gauge (\ref{wzgauge}), that is, $\delta_\xi E_\mu^{\ A}=0=\delta_\xi E^{\dot{\mu}}_{\ A}$. Thus, we no longer consider components $M=\mu, \dot{\mu}$. 

Evaluating the $M=m$ components of (\ref{newkilling}) at $\theta=0=\bar{\theta}$, we get
\begin{subequations}\label{tetradkill}
\begin{align}
-\mathscr{D}_m \xi^a-\xi^b T_{bm}^{\ \ a}|+e_m^{\ b} \xi^c \omega_{cb}^{\ \ a}&=e_m^{\ b} L_b^{\ a}|, \\
-\xi^b T_{bm}^{\ \ \alpha}|+\tfrac{1}{2} \psi_m^{\ \beta} \xi^c \omega_{c \beta}^{\ \ \alpha}&=\tfrac{1}{2} \psi_m^{\ \beta} L_\beta^{\ \alpha}|. \label{spinorb}
\end{align}
\end{subequations}
The equation with index $\dot{\alpha}$ follows by complex conjugation of (\ref{spinorb}). The required torsion components are computed in (\ref{torsions}), substituting them into (\ref{tetradkill}), we get
\begin{subequations}\label{tetradkill2}
\begin{align}
-\mathscr{D}_m \xi^a+e_m^{\ b} \xi^c \omega_{cb}^{\ \ a}&=e_m^{\ b} L_b^{\ a}, \label{tetradeq}\\
-\xi^b \psi_{bm}^{\ \ \alpha}-\tfrac{i}{6} e_m^{\ c} \xi^b ( (\psi_b\sigma^d \bar{\sigma}_c)^\alpha b_d+3 \psi_b^{\ \alpha} b_c-\bar{\psi}_{b \dot{\gamma}} \bar{\sigma}_c^{\dot{\gamma} \alpha} M)+\psi_m^{\ \beta} \xi^c \omega_{c \beta}^{\ \ \alpha}&=\psi_m^{\ \beta} L_\beta^{\ \alpha}. \label{spinor}
\end{align}
\end{subequations}

It turns out that more simple expressions can be given if we consider non-vanishing local spinor parameters in (\ref{prev}). First of all, we switch to spacetime index by substituting $\xi^a=\xi^l e_l^{\ a}$ in (\ref{tetradkill2}). After some manipulations detailed in (\ref{auxeq}), we obtain
\begin{subequations}\label{firstkilling}
	\begin{align}
		-\xi^l \partial_l e_m^{\ a}-e_l^{\ a} \partial_m \xi^l+\tfrac{i}{2} \xi^l (\psi_l \sigma^a \bar{\psi}_m-\psi_m \sigma^a \bar{\psi}_l)&=e_m^{\ b} L_b^{\ a},  \label{tetradR} \\
		-\xi^l \partial_l \psi_m^{\ \alpha}-(\partial_m \xi^l) \psi_l^{\ \alpha}+\mathcal{D}_m (\xi^l \psi_l^{\ \alpha})-\tfrac{i}{6} e_m^{\ b} (\xi^l (\psi_l \sigma^c \bar{\sigma}_b)^\alpha b_c- \xi^l (\bar{\psi}_l \bar{\sigma}_b)^\alpha M+3 \xi^l \psi_l^{\ \alpha} b_b)&=\psi_m^{\ \beta} L_\beta^{\ \alpha}. \label{spinorR}
	\end{align}
\end{subequations}

The expressions for the auxiliary fields, derived in Appendix \ref{appc}, are
\begin{subequations}\label{firstaux}
\begin{align}
-\xi^l \partial_l M+\xi^l \psi_l^{\ \alpha} (\sigma^{ab})_\alpha^{\ \beta} \psi_{ab \beta}-\tfrac{i}{2} \xi^l (\psi_l \sigma^a \bar{\psi}_a) M+\tfrac{i}{2} \xi^l \psi_l^{\ \alpha} \psi_{a \alpha} b^a&=0, \label{newscal} \\
-\xi^l \partial_l b_a-\tfrac{3}{2} \xi^l \psi_l^{\ \delta} \mathscr{D}_\delta G_a|-\tfrac{3}{2} \xi^l \bar{\psi}_{l \dot{\delta}} \bar{\mathscr{D}}^{\dot{\delta}} G_a|&=b_b L_b^{\ a}. \label{newvector}
\end{align}
\end{subequations}

Taking into account the supergravity transformations (\ref{sugratrans}), one notices that the left-hand sides of (\ref{firstkilling}) and (\ref{firstaux}) involve a field-dependent supergravity transformation parameterized by  $\tilde{\zeta}^{\alpha}=-\frac{1}{2} \xi^m \psi_m^{\ \alpha}$. That is, the local transformation (\ref{prev}) induces a spacetime diffeomorphism and a field dependent supergravity transformation,
\begin{align}
\delta_{\xi^a} A=\delta_{\xi^m} A+\delta_{\tilde{\zeta}^\alpha} A.
\end{align}

Therefore, we cancel the contributions $\delta_{\tilde{\zeta}} A$ by including  non-vanishing spinor transformation parameters, 
\begin{align}\label{diffeo}
\xi^A|=(\xi^a, \xi^\alpha,\bar{\xi}_{\dot{\alpha}})=(\xi^a, \tfrac{1}{2} \xi^b \psi_b^{\ \alpha},\tfrac{1}{2} \xi^b \bar{\psi}_{b \dot{\alpha}}),
\end{align}

According to (\ref{local}), (\ref{diffeo}) correspond to a pure coordinate transformation $\xi^M|=(\xi^m(x), \xi^\mu(x),\bar{\xi}_{\dot{\mu}}(x))=(\xi^b e_b^{\ m},0,0)$. Unlike the local supergravity transformation parameters $\zeta^\alpha$ \cite{wessbagger}, we choose the spacetime parameters $\xi^m(x)$ field-independent, so that they do not transform under supergravity.

\section{Killing equations of supergravity}\label{sec4}
Defining isometries as transformations of the form (\ref{diffeo}), then our prospect supergravity Killing equations read
\begin{subequations}\label{sugrakilling}
\begin{align}
-\xi^l \partial_l e_m^{\ a}-e_l^{\ a} \partial_m \xi^l&=e_m^{\ b} L_b^{\ a}, \label{tetradeqf}\\
-\xi^l \partial_l \psi_m^{\ \alpha}-\psi_l^{\ \alpha} \partial_m \xi^l&=\psi_m^{\ \beta} L_\beta^{\ \alpha} \label{spinorvector}\\
\xi^{l} \partial_l M&=0, \label{scalar} \\
-\xi^l \partial_l b_a&=b_b L^b_{\ a}.\label{vector}
\end{align}
\end{subequations}
This set contains the tetrad Killing equations (\ref{killingtetrad}) and additional equations for the other components of the supergravity multiplet. As mentioned previously, (\ref{tetradeqf}) amounts to 10 equations only, plus 6 equations of the auxiliary fields (\ref{scalar}), (\ref{vector}), we have 16 equations for the bosons, on par with the 8 complex equations contained in (\ref{spinorvector}).

The input of (\ref{sugrakilling}) is a given supergravity multiplet, ideally a solution of the field equations, while the output is a set of linearly independent vector fields and their corresponding infinitesimal Lorentz generators. Now, since we are interested in extending symmetric configurations in gravity to supergravity, we are using (\ref{sugrakilling}) the other way around. The strategy is to take the tetrad and Killing vectors of a given spacetime, solve for the Lorentz generators from  (\ref{tetradeqf}), and use the remaining equations in (\ref{sugrakilling}) to determine the form of the spinor-vector and auxiliary fields.

On the other hand, the preservation of isometries under supergravity transformations (\ref{sugratrans}) imposes additional  requirements. Replacing  $e_m^{\ a}$ by $\delta_\zeta e_m^{\ a}$ in (\ref{tetradeqf}), and using (\ref{spinorvector}) and (\ref{generators}), we get
\begin{align*}
-i \xi^l \partial_l (\psi_m \sigma^a \bar{\zeta})-i (\psi_l \sigma^a \bar{\zeta}) \partial_m \xi^l+h.c.&=-i (\xi^l \partial_l \psi_m+\psi_l \partial_m \xi^l) \sigma^a \bar{\zeta}-i \psi_m \sigma^a \xi^l \partial_l \bar{\zeta}+h.c. \nonumber \\
&=i \psi_m^{\ \beta} L_\beta^{\ \alpha} \sigma^a_{\alpha \dot{\alpha}} \bar{\zeta}^{\dot{\alpha}}-i \psi_m^{\ \alpha} \sigma^a_{\alpha \dot{\alpha}} (\xi^l \partial_l \bar{\zeta})^{\dot{\alpha}}+h.c., \\
&=i (\psi_m \sigma^b \bar{\zeta}-\zeta \sigma^a \bar{\psi}_m) L_b^{\ a} \\
&=i \psi_m^{\ \beta} L_\beta^{\ \alpha} \sigma^a_{\alpha \dot{\alpha}} \bar{\zeta}^{\dot{\alpha}}+i \psi_m^{\ \alpha} \sigma^a_{\alpha \dot{\alpha}} \bar{\zeta}^{\dot{\beta}} L_{\dot{\beta}}^{\ \dot{\alpha}}+h.c.
\end{align*}
Therefore, supergravity transformations respecting the symmetries satisfy 
\begin{align}\label{sugrapar}
	-\xi^l \partial_l \zeta^\alpha(x)=\zeta^\beta(x) L_\beta^{\ \alpha}(x).
\end{align}
Similarly, replacing $\psi_m^{\ \alpha}$ by $\delta_\zeta \psi_m^{\ \alpha}$ in (\ref{spinorvector}), yields
\begin{align}\label{sugracon}
\delta_\xi \omega_{m \beta}^{\ \ \ \alpha}\equiv -\xi^l \partial_l \omega_{m \beta}^{\ \ \ \alpha}-\omega_{l \beta}^{\ \ \alpha} \partial_m \xi^l=\omega_{m \beta}^{\ \ \ \gamma} L_\gamma^{\ \alpha}-\partial_m L_\beta^{\ \alpha}-L_\beta^{\ \gamma} \omega_{m \gamma}^{\ \ \ \alpha}.
\end{align}

As an example let's consider the isometries of spatially flat homogeneous and isotropic or $k=0$ FRW, spacetime \cite{ellis}. In this case, the Killing vectors associated to homogeneity have vanishing structure constants, thus, there exists spatial coordinates $x^i$ such that $T_i=\partial_i$. Clearly, a particular solution of (\ref{sugrakilling}) is a supergravity multiplet depending only on time $x^0=t$,
\begin{align}\label{homsugra}
e_m^{\ a}=e_m^{\ a}(t), && \psi_m^{\ \alpha}=\psi_m^{\ \alpha}(t), && b_a=b_a(t), && M=M(t),
\end{align}
for which $L_b^{\ a}(\partial_i)=0$. Spatial dependence, leading to non-vanishing $L_b^{\ a}(\partial_i)$, may enter through a local Lorentz transformation. Condition (\ref{sugrapar}) is satisfied with a $\xi^\alpha=\xi^{\alpha}(t)$. 

Now, spatial isotropy comprises three spacelike Killing vectors satisfying the $\frak{so}(3)$ algebra, whose orbits are 2-dimensional spheres \cite{ryanbook,taub}. In flat space, they are the ordinary rotation generators, $J_i=\epsilon_{ij}^{\ \ k} x^j \partial_k$. Choosing the following tetrad
\begin{align}\label{tetradfrw}
\bm e^{\tilde{0}}=dt\ N(t), && \bm e^{\tilde{i}}=dx^i\ a(t),
\end{align}
where tildes are used for Lorentz indices, then (\ref{solvelorentz}) yields $L_b^{\ a}(J_i)=-\epsilon_{ij}^{\ \ k} e_b^{\ m} e_k^{\ a} \delta_m^j=-\epsilon_{ij}^{\ \ k} \delta_b^{\ j} \delta_k^{\ a}$. 

Now, the $m=0$ and $m=j$ components of (\ref{spinorvector}) read, respectively,
\begin{align}\label{rotspinor}
0=\psi_0^{\ \beta}(t) L_\beta^{\ \alpha}(J_i), && \psi_k^{\ \alpha}(t) \epsilon_{ij}^{\ \ k}=\psi_j^{\ \beta}(t) L_\beta^{\ \alpha}(J_i),
\end{align}
with the spinor generators $L_\beta^{\ \alpha}(J_i)$ (\ref{generators}) proportional to the Pauli matrices. Thus, (\ref{rotspinor}) yield a vanishing spinor-vector. 

Thus, the answer to the question raised in Section \ref{sec3} is that, adopting (\ref{sugrakilling}) as the supergravity Killing equations, spatial isotropy implies a vanishing Rarita-Schwinger field. To define a nontrivial FRW background of supergravity, we consider a set of less stringent equations in place of (\ref{spinorvector}).

\subsection{Relaxed spinor-vector equations}\label{sec5}
As mentioned in Section \ref{sec2}, as far as the Killing vectors concern, equations (\ref{killingtetrad}) are equivalent to (\ref{killingmetric}), the extra equations determining the Lorentz generators. Thus, we look for equations for the spinor-vector that do not depend on the frame of reference. 

We proceed as with the tetrad, contracting (\ref{spinorvector}) with $\psi_{n\alpha}$ and symmetrizing in $m, n$. Using the anti-commutativity of spinors and (\ref{not1}), we have
\begin{align*}
(\delta_\xi \psi_m^{\ \alpha}) \psi_{n\alpha}+(\delta_\xi \psi_n^{\ \alpha}) e_{m\alpha}&=-\xi^l (\partial_l \psi_m^{\ \alpha}) \psi_{n \alpha}-\psi_l^{\ \alpha} (\partial_m \xi^l) \psi_{n \alpha}-\xi^l (\partial_l \psi_n^{\ \alpha}) \psi_{m \alpha}-\psi_l^{\ \alpha} (\partial_n \xi^l) \psi_{m \alpha} \\
&=-\xi^l (\partial_l \psi_m^{\ \alpha}) \psi_{n \alpha}-\xi^l \psi_m^{\ \alpha} (\partial_l \psi_{n \alpha})-\psi_l^{\ \alpha} \psi_{n \alpha} (\partial_m \xi^l)-\psi_m^{\  \alpha} \psi_{l \alpha} (\partial_n \xi^l) \\
&=\psi_m^{\ \beta} L_\beta^{\ \alpha} \psi_{n \alpha}+\psi_n^{\ \beta} L_\beta^{\ \alpha} \psi_{m \alpha}
=-\psi_m^{\ \alpha} L_{\alpha \beta} \psi_n^{\ \beta}+\psi_m^{\ \alpha} L_{\beta \alpha} \psi_n^{\ \beta}.
\end{align*}
Therefore, since $L_{\alpha \beta}=L_{\beta \alpha}$, we obtain
\begin{align}\label{spinormetric}
	\xi^l \partial_l (\psi_m^{\ \alpha} \psi_{n \alpha})+\psi_l^{\ \alpha} \psi_{n \alpha} \partial_m \xi^l+\psi_m^{\ \alpha} \psi_{l \alpha} \partial_n \xi^l=0,
\end{align}
which is our replacement for (\ref{spinorvector}). The quantities $\psi_m^{\ \alpha} \psi_{n\alpha}=\psi_n^{\ \alpha} \psi_{m\alpha}$ are suitable to define a complex ten-parameter equivalence class of the spinor-vectors related by Lorentz transformation, analogous to the class of tetrads of Section \ref{sec2}. 

Since there are no multiplicative inverse of spinors, (\ref{spinormetric}) does not imply (\ref{spinorvector}) but only the following contracted form
\begin{align}
	(-\xi^l \partial_l \psi_m^{\ \alpha}-\psi_l^{\ \alpha} \partial_m \xi^l) \psi_{n \alpha}=\psi_m^{\ \gamma} L_\gamma^{\ \alpha} \psi_{n\alpha}.
\end{align}
That its right-hand side vanishes identically for $n=m$ allows a nonzero $\psi_0$, in contrast to (\ref{rotspinor}).

Let's consider the FRW symmetries once again. For the rotation generators and the homogeneous spinor-vector $\psi_m^{\ \alpha}(t)$, (\ref{spinormetric}) yields the following constraints,
\begin{align}\label{spinormetriccons}
	\psi_a \psi_b=0,\ (a\ne b), && \psi_1 \psi_1=\psi_2 \psi_2=\psi_3 \psi_3.
\end{align}
If none of the $\psi_m$ vanishes, the general solution of (\ref{spinormetriccons}) can be expressed in terms of a single Weyl spinor $\psi^\alpha$, as follows
\begin{align}\label{three}
	\psi_0=c_0 (t) \begin{bmatrix}
		\psi^{1} \\
		\psi^{2}
	\end{bmatrix}, && 
	\psi_1=c_1(t) \begin{bmatrix}
		\psi^{2} \\
		\psi^{1}
	\end{bmatrix}, &&
	\psi_2=c_1(t) \begin{bmatrix}
		-i \psi^{2} \\
		i \psi^{1}
	\end{bmatrix}, &&
	\psi_3=c_1(t) \begin{bmatrix}
		\psi^{1} \\
		-\psi^{2}
	\end{bmatrix}
\end{align}
where $c_0(t)$ and $c_1(t)$ are real functions\footnote{Complex functions do not yield a different solution.}. (\ref{three}) satisfies (\ref{spinormetriccons}) with an additional property $c_0^2 \psi_i \psi_i=c_1^2 \psi_0 \psi_0$. Therefore, equations (\ref{spinormetric}) restrict the number of real independent components in the fermionic sector to five, contained in $\psi$ and the ratio $c_0(t)/c_1(t)$. In the bosonic sector we have the lapse $N$ and the scale factor, plus the three components of $M, M^*$ and $b^0$, making a total of five components too. However, we have not yet consider supergravity transformations.

Isometries of the metric are preserved by supergravity if
\begin{align}\label{new0}
\delta_\xi (\delta_\zeta e_m^{\ a})\equiv -\xi^l \partial_l (\delta_\zeta e_m^{\ a})-\delta_\zeta e_l^{\ a} \partial_m \xi^l=(\delta_\zeta e_m^{\ b}) L_b^{\ a}+e_m^{\ b} K_b^{\ a} 
\end{align}
where $L_b^{\ a}$ is given in (\ref{solvelorentz}) and $K_{ba}=-K_{ab}$ is another Lorentz generator. Indeed, if (\ref{new0}) holds, then
\begin{align*}
\delta_\xi (\delta_\zeta e_m^{\ a}) e_{na}+e_{ma} \delta_\xi (\delta_\zeta e_n^{\ a})&=(\delta_\zeta e_m^{\ b}) L_b^{\ a} e_{na}+(\delta_\zeta e_n^{\ b}) L_b^{\ a} e_{ma}+e_m^{\ b} K_b^{\ a} e_{na}+e_n^{\ b} K_b^{\ a} e_{ma} \nonumber \\
&=-(\delta_\zeta e_m^{\ b})  e_{na} L^a_{\ b}-(\delta_\zeta e_{n b}) e_m^{\ a} L_a^{\ b}+e_m^{\ b} K_{ba} e_n^{\ a}+e_n^{\ a} K_{ab} e_m^{\ b} \nonumber \\
&=-(\delta_\zeta e_m^{\ b}) \delta_\xi e_{nb}-(\delta_\xi e_m^{\ b}) (\delta_\zeta e_{n b}),
\end{align*}
using (\ref{tetradeqf}) and $K_{ab}=-K_{ba}$. Thus, (\ref{new0}) implies that 
\begin{align}
\delta_\xi [(\delta_\zeta e_m^{\ a}) e_{na}+e_m^{\ a}\delta_\zeta e_{na}]=\delta_\xi [\delta_\zeta (e_m^{\ a} e_{na})]=0.
\end{align}

\subsection{A spin-$\frac{1}{2}$ solution}
In cases such as the diagonal FRW metric, we require that 
\begin{align}\label{deltametric}
\delta_\zeta (e_m^{\ a} e_{na})\propto e_m^{\ a} e_{na}.
\end{align}
Using the product rule and the transformation of the tetrad (\ref{sugratetrad}), we see that the left-hand side of (\ref{deltametric}) is linear in the spinor parameters $\zeta^\alpha$, $\bar{\zeta}^{\dot{\alpha}}$. Thus, we write 
\begin{align}\label{preserve}
i (\psi_m \sigma^a \bar{\zeta}-\zeta \sigma^a \bar{\psi}_m) e_{na}+i e_{ma} (\psi_n \sigma^a \bar{\zeta}-\zeta \sigma^a \bar{\psi}_n)=i e_m^{\ a} e_{na} (\lambda \zeta-\bar{\lambda} \bar{\zeta}).
\end{align}
for some undetermined spinor $\lambda$ (see conventions (\ref{not1}) and (\ref{not2})).

Equation (\ref{preserve}) comprises the sum of complex conjugate contributions and may be satisfied in several ways. We pursue here the case in which both components vanish independently, that is
\begin{align}\label{vanish}
i \psi_m \sigma^a \bar{\zeta} e_{na}+i e_{m a} \psi_n \sigma^a \bar{\zeta}+\tfrac{i}{2} (e_m^{\ a} e_{na}+e_n^{\ a} e_{ma}) \bar{\lambda} \bar{\zeta}=0.
\end{align}
This shows that
\begin{align}
S_{mn}\equiv i \psi_m \sigma^a \bar{\zeta} e_{n a}+\tfrac{i}{2} e_m^{\ a} e_{na} \bar{\lambda} \bar{\zeta}=-S_{nm}.
\end{align}
thus, taking the trace of $S_{mn}$, we get
\begin{align}\label{trace}
i \psi_a \sigma^a \bar{\zeta}+2 i \bar{\lambda} \bar{\zeta}=0.
\end{align}

Now, substituting into (\ref{trace}) the spin-$\frac{1}{2}$ and spin-$\frac{3}{2}$ decomposition of the spinor-vector \cite{wessbagger},
\begin{align}\label{decomposed}
	\psi_a^{\ \alpha}=-\tfrac{1}{2} \bar{\sigma}_a^{\dot{\beta} \beta} \psi_{\beta \ \dot{\beta}}^{\ \alpha}=-\tfrac{1}{2} \bar{\sigma}_a^{\dot{\beta} \beta} \epsilon^{\alpha \delta} (\epsilon_{\beta \delta} \bar{\psi}_{\dot{\beta}}+W_{\beta \delta \dot{\beta}}),
\end{align}
where $W_{\beta \delta \dot{\beta}}=W_{(\beta \delta) \dot{\beta}}$, yields
\begin{align}
	\bar{\lambda}_{\dot{\alpha}}=\bar{\psi}_{\dot{\alpha}}.
\end{align}
Moreover, going back to (\ref{vanish}), we find that $W_{\alpha \delta \dot{\alpha}}=0$. Therefore, (\ref{deltametric}) implies a spin-$\tfrac{1}{2}$ truncation of the Rarita-Schwinger field
\begin{equation}\label{anzats}
\psi_m^{\ \alpha}(x)=\tfrac{1}{2} e_m^{\ a}(x) \bar{\psi}_{\dot{\beta}}(x) \bar{\sigma}_a^{\dot{\beta} \alpha},
\end{equation}

Now, let's consider (\ref{spinormetric}). First of all,
\begin{align}\label{psimn}
\psi_m^{\ \alpha} \psi_{n\alpha}&=\tfrac{1}{4} e_m^{\ a} e_n^{\ b} \bar{\psi}_{\dot \alpha} \bar{\sigma}_a^{\dot{\alpha} \alpha} \epsilon_{\alpha \beta} \bar{\psi}_{\dot{\beta}} \bar{\sigma}_b^{\dot{\beta} \beta}=-\tfrac{1}{8} e_m^{\ a} e_n^{\ b} \bar{\psi}_{\dot{\gamma}} \bar{\psi}^{\dot{\gamma}} \bar{\sigma}_a^{\dot{\alpha} \alpha} \epsilon_{\dot \alpha \dot{\beta}} \epsilon_{\alpha \beta} \bar{\sigma}_b^{\dot{\beta} \beta} \nonumber \\
&=-\tfrac{1}{8} e_m^{\ a} e_n^{\ b} \bar{\psi}_{\dot{\gamma}} \bar{\psi}^{\dot{\gamma}} \bar{\sigma}_a^{\dot{\alpha} \alpha} \sigma_{b \alpha \dot{\alpha}}=\tfrac{1}{4} e_m^{\ a} e_{na} \bar{\psi}_{\dot{\alpha}} \bar{\psi}^{\dot{\alpha}}.
\end{align}
Clearly, (\ref{psimn}) satisfies the Killing equation (\ref{spinormetric}) if the tetrad $e_m^{\ a}(x)$ satisfies (\ref{tetradeqf}) and $\bar{\psi}_{\dot{\alpha}} \bar{\psi}^{\dot{\alpha}}$ is an invariant,
\begin{equation}\label{spinorcalar}
\xi^l \partial_l (\bar{\psi} \bar{\psi})=0.
\end{equation}

On the other hand, substituting (\ref{anzats}) into the left-hand side of (\ref{new0}) and rearranging, we get
\begin{align}
K_{ba}=i \bar{\psi} (L_b^{\ d} \bar{\sigma}_d \sigma_a-L_a^{\ d} \bar{\sigma}_d \sigma_b) \bar{\zeta}+2 i L_{ba}\bar{\psi} \bar{\zeta}-2 i \xi^l \partial_l (\bar{\psi} \bar{\sigma}_{ba} \bar{\zeta})+i \eta_{ba} \xi^l \partial_l (\bar{\psi} \bar{\zeta})+h.c.,
\end{align}
which is antisymmetric if the coefficient of $\eta_{ba}$ vanishes. Therefore, isometries of the metric are preserved by supergravity transformations provided that
\begin{align}\label{invar}
\xi^l \partial_l [i (\bar{\psi} \bar{\zeta}-\zeta \psi)]=0.
\end{align}
(\ref{invar}) is the replacement of (\ref{sugrapar}) for the case of (\ref{anzats}). This conclusion also follows from 
\begin{align}\label{deltametric2}
\delta_\zeta (e_m^{\ a} e_{na})=-2 i e_m^{\ a} e_{n a} (\bar{\psi} \bar{\zeta}-\zeta \psi).
\end{align}

On the other hand, from (\ref{psimn}),
\begin{align}\label{deltametpsi}
\delta_\zeta (\psi_m \psi_n)=\tfrac{1}{4} \delta_\zeta (e_m^{\ a} e_{na}) \bar{\psi}_{\dot{\alpha}} \bar{\psi}^{\dot{\alpha}}+\tfrac{1}{2} e_m^{\ a} e_{na} (\delta_\zeta \bar{\psi}_{\dot{\alpha}}) \bar{\psi}^{\dot{\alpha}}.
\end{align}

The transformation of the single spinor $\bar{\psi}$ follows from (\ref{anzats}). Using (\ref{sugratetrad}) and (\ref{sugraspinor}), we get
\begin{align}\label{deltapsi}
\delta_\zeta \bar{\psi}_{\dot{\delta}}&=e_b^{\ m} [-\tfrac{1}{2} (\delta_\zeta \psi_m^{\ \alpha})+\tfrac{1}{4} (\delta_\zeta e_m^{\ a}) \bar{\psi}_{\dot{\gamma}} \bar{\sigma}^{\dot \gamma \alpha}_a] \sigma^b_{\alpha \dot{\delta}} \nonumber \\
&=((\mathscr{D}_m \zeta) \sigma^m)_{\dot{\delta}}-\tfrac{2i}{3} M \bar{\zeta}_{\dot{\delta}}+\tfrac{i}{6} b_d (\zeta \sigma^d)_{\dot{\delta}}-\tfrac{i}{2} (\zeta \psi+2 \bar{\psi} \bar{\zeta}) \bar{\psi}_{\dot{\delta}}
\end{align}
where $\sigma^m\equiv \sigma^b e_b^{\ m}$. Thus, substituting (\ref{deltametric2}) and (\ref{deltapsi}) into (\ref{deltametpsi}) and simplifying, we get
\begin{align}
\delta_\zeta (\psi_m \psi_n)=\tfrac{1}{2} e_m^{\ a} e_{na} [(\mathscr{D}_l \zeta) \sigma^l \bar{\psi}-\tfrac{2i}{3} M \bar{\zeta} \bar{\psi}+\tfrac{i}{6} b_d \zeta \sigma^d \bar{\psi}+\tfrac{i}{2} \zeta \psi \bar{\psi} \bar{\psi}]
\end{align}
Thus, the term in square brackets must also be an invariant under the isometries.

As an example, for k=0 FRW cosmology, (\ref{anzats}) yields 
\begin{align}
	\psi_t&=\tfrac{1}{2} N(t) \begin{bmatrix}
		\bar{\psi}_{\dot{1}} \\
		\bar{\psi}_{\dot{2}}
	\end{bmatrix}, && \psi_r=\tfrac{1}{2} a(t) \begin{bmatrix}
		\bar{\psi}_{\dot{2}} \\
		\bar{\psi}_{\dot{1}}
	\end{bmatrix}, \\
	\psi_\theta&=\tfrac{1}{2} a(t) \begin{bmatrix}
		-i \bar{\psi}_{\dot{2}} \\
		i \bar{\psi}_{\dot{1}}
	\end{bmatrix}, && 
	\psi_\phi=\tfrac{1}{2} a(t) \begin{bmatrix}
		-\bar{\psi}_{\dot{1}} \\
		\bar{\psi}_{\dot{2}}
	\end{bmatrix},
\end{align}
with $\bar{\psi}_{\dot{\alpha}}=\bar{\psi}_{\dot{\alpha}}(t)$ and $\zeta^\alpha=\zeta^{\alpha}(t)$. Spatial dependence can be introduced with a local Lorentz transformation, while still satisfying (\ref{spinormetric}).

Thus, in order to preserve the diagonal form of the FRW metric, we have to set $c_0=N(t)$ and $c_1=a(t)$, which leaves just the four real components of the Weyl spinor $\psi$. To preserve the supersymmetry a boson component is also eliminated, from  (\ref{deltametric2}) we get, in particular, $\delta_\zeta N/N=\delta_\zeta a/a$, or $\delta_\zeta \ln N=\delta_\zeta \ln a$. This can be put as the time gauge being fixed automatically to conformal time $\eta$ corresponding to $N=a$.

The supergravity Lagrangian corresponding to the particular solution (\ref{anzats}) takes the form
\begin{align}
	\mathcal{L}_{SG}&=-\tfrac{1}{2} e R-\tfrac{1}{3} e M^* M+\tfrac{1}{3} e b^a b_a+\tfrac{1}{2} e \epsilon^{klmn} (\bar{\psi}_k \bar{\sigma}_l \tilde{\mathcal{D}}_m \psi_n-\psi_k \sigma_l \tilde{\mathcal{D}}_m \bar{\psi}_n) \nonumber \\
	&\to -\tfrac{1}{2} e \tilde{R}-\tfrac{3}{4} i e e_b^{\ m} ((\partial_m \bar{\psi}) \bar{\sigma}^b \psi-\bar{\psi} \bar{\sigma}^b \partial_m \psi)-\tfrac{3}{8} e \epsilon^{bcmn} \bar{\psi} \bar{\sigma}_c \psi \partial_m e_{nb}+\tfrac{3}{32} e \psi \psi \bar{\psi} \bar{\psi}
\end{align}
where $\tilde{R}$ is the bosonic curvature and the auxiliary field have been integrated out. In the FRW example $\mathcal{L}^{FRW}=\mathcal{L}^{FRW}(a,\psi)$. The gauge fixing can be avoided with different solutions, for example, one such that 
\begin{align}
\bar{\psi}_{\dot{\alpha}} e_i^{\dot{\alpha} \alpha} \psi_{0\alpha}=0, && \psi_0^{\ \alpha} e_{i \alpha \dot{\alpha}} \bar{\zeta}^{\dot{\alpha}}=0=\psi_i^{\ \alpha} \sigma_{0\alpha \dot{\alpha}} \bar{\zeta}^{\dot{\alpha}}.
\end{align}
where $e_{i \alpha \dot{\alpha}}\equiv e_i^{\ a} \sigma_{a\alpha \dot{\alpha}}$, etc., similar to those required in other approaches e.g., \cite{moniz}. With the present solution, the spinors $\psi^\alpha(t)$ and $\zeta^\alpha(t)$ are not subjected to further constraints and we have not fixed the frame of reference.

\section{Conclusions}\label{sec6}
In this work, we consider an extension of the Killing equations for N=1 4D supergravity using a superspace approach. First, the Killing equations in the language of tetrads were promoted to superfield equations for the vielbein and evaluated in the Wess-Zumino gauge. This leads to a first set of prospect supergravity Killing equations whose geometric content can be stated as the invariance of the supergravity multiplet under an isometry, up to a Lorentz transformation. These equations are reinforced by consideration of supergravity transformations, which leads to the same kind of invariance condition on the spinor parameters.

For spatial isotropy, the spinor-vector equations imply a vanishing Rarita-Schwinger field. Thus, for a classical isotropic configuration with non-vanishing spinor-vector, we have to relax the equations. Taking into account that the admission of Killing vectors is not a property of a particular tetrad, but of the whole equivalence class of tetrads determined by a metric, an alternative set of equations is derived by contraction of the local spinor index of the Rarita-Schwinger field. This yields the ordinary Killing equations for the symmetric tensor $\psi_m \psi_n$. We describe a particular solution of the relaxed equations consisting of a spin-$\frac{1}{2}$ truncation of the Rarita-Schwinger field which works, in particular, for the FRW symmetries with positive, negative or zero spatial curvature. Thus, while the first linear equations kill the spinor-vector altogether for spatial isotropy, the alternative quadratic equations allow a spin-$\tfrac{1}{2}$ field, that is, an FRW supermultiplet consisting of the scalar field $a(t)$ and a spin-$\frac{1}{2}$ field $\psi$.

More general definitions of isometries that involve invariance of the supergravity multiplet not only up to a Lorentz transformation, but also up to local coordinate and supergravity transformations, as done with the k=1 FRW model in \cite{death88,death92,asano}, will be considered in future work. There are several aspects that have yet to be addressed such as the isolation of the symmetry reduced supergravity multiplet and the effect of symmetries on the field equations. Other topics  are the formulation of Killing equations for higher-N supergravities and the study of quantum cosmological models constructed using this approach.

\appendix
\section{Further details}\label{app1}
\subsection{Notation and useful results}\label{appa}
Local spinor indices can be raised or lowered according to $X^\alpha=\epsilon^{\alpha \beta} X_\beta$, $X_\alpha=\epsilon_{\alpha \beta} X^\beta$, where $\epsilon^{12}=1=\epsilon_{21}$ and $\epsilon_{12}=-1=\epsilon^{21}$, and similarly for dotted indices. Then, using the anti-commutativity of individual spinor components, we have
\begin{align}\label{not1}
	\psi \zeta\equiv \psi^\alpha \zeta_\alpha=\zeta^\alpha \psi_\alpha \equiv \zeta \psi, && \bar{\psi} \bar{\zeta}\equiv \bar{\psi}_{\dot{\alpha}} \bar{\zeta}^{\dot{\alpha}}=\bar{\zeta}_{\dot{\alpha}} \bar{\psi}^{\dot{\alpha}}\equiv \bar{\zeta} \bar{\psi}.
\end{align}

Lorentz generators are related by
\begin{align}\label{generators}
L_\alpha^{\ \beta}(x)=-\tfrac{1}{2} (\sigma^{ab})_\alpha^{\ \beta} L_{ab}(x), && L^{\dot{\alpha}}_{\ \dot{\beta}}=-\tfrac{1}{2} (\bar{\sigma}^{ab})^{\dot{\alpha}}_{\ \dot{\beta}} L_{ab}
\end{align}
where $\sigma^{ab}=\frac{1}{4} (\sigma^a \bar{\sigma}^b-\sigma^b \bar{\sigma}^a)$ ($\bar{\sigma}^{ab}=\frac{1}{4} (\bar \sigma^a \sigma^b-\bar \sigma^b \sigma^a)$). The Infeld–Van der Waerden symbols have index structure $\sigma^a_{\alpha \beta}$ and $\bar{\sigma}^{a\dot{\beta} \alpha}\equiv \epsilon^{\dot{\beta} \dot{\gamma}} \epsilon^{\alpha \delta} \sigma^a_{\delta \dot{\gamma}}$. We made use of the following contractions
\begin{align}\label{not2}
	\psi \sigma^a \zeta\equiv \psi^\alpha \sigma^a_{\alpha \dot{\beta}} \bar{\zeta}^{\dot{\beta}}, && \bar{\psi} \bar{\sigma}^a \zeta\equiv \bar{\psi}_{\dot{\alpha}} \bar{\sigma}^{a \dot{\alpha} \beta} \zeta_\beta.
\end{align}
$\sigma^0\equiv -\bm 1=-\bar{\sigma}^0$ and $\sigma^i=-\bar{\sigma}^i$ are the standard Pauli matrices.

On the other hand,  covariant derivatives are defined with a Lie algebra-valued connection, $\phi_{MBA}(z)=-(-)^{ab} \bm \phi_{MBA}(z)$ (exponent $a$ takes value $0$ if $A=a$ or $1$ if $A=\alpha, \dot{\alpha}$), as follows
\begin{align}\label{covder}
\mathscr{D}_M V^A=\partial_M V^A+(-)^{mb} V^B \phi_{MB}^{\ \ \ \ A}, && \mathscr{D}_M V_A=\partial_M V_A-\phi_{MAB} V^B
\end{align}
(exponent $m$ takes value $0$ if $M=m$ or $1$ if $M=\mu, \dot{\mu}$).

The inverse vielbein superfields $E_A^{\ M}(z)$ satisfy $E_M^{\ \ A} E_A^{\ N}=\delta_M^{\ N}$, $E_A^{\ M} E_M^{\ B}=\delta_A^{\ B}$. In the W-Z gauge, they are given by
\begin{align}\label{inverse}
E_A^{\ M}(z)|=\begin{bmatrix}
e_a^{\ m}(x) & -\tfrac{1}{2} \psi_a^{\ \mu}(x) & -\tfrac{1}{2} \bar{\psi}_{a\dot{\mu}}(x) \\
0 & \delta_\alpha^{\ \mu} & 0 \\
0 & 0 & \delta^{\dot{\alpha}}_{\ \dot{\mu}}
\end{bmatrix},
\end{align}
where $\psi_a^{\ \alpha}\equiv e_a^{\ n} \psi_n^{\ \alpha}$, $\psi_a^{\ \mu}=e_a^{\ l} \psi_l^{\ \beta} \delta_\beta^{\ \mu}$, and
\begin{align}\label{doble}
\psi_{mn}^{\ \ \alpha}\equiv \partial_m \psi_n^{\ \alpha}+\psi_n^{\ \beta} \omega_{m\beta}^{\ \ \alpha}-\partial_n \psi_m^{\ \alpha}-\psi_m^{\ \beta} \omega_{n \beta}^{\ \ \alpha}
\end{align}

The torsion components are given by
\begin{align}\label{torsion}
	T_{NM}^{\ \ \ A}(z)=\partial_N E_M^{\ A}-(-)^{nm} \partial_M E_N^{\ A}+(-)^{n(b+m)} E_M^{\ B} \phi_{NB}^{\ \ \ A}-(-)^{mb} E_N^{\ B} \phi_{MB}^{\ \ \ A}.
\end{align}  
They are subjected to the Bianchi identities and to the following set of covariant constraints  
\begin{align}\label{const}
\begin{split}
	T_{\underline{\alpha} \underline{\beta}}^{\ \ \ \underline{\gamma}}=0, \ \ \ \ \ T_{\alpha \beta}^{\ \ c}=0=T_{\dot{\alpha} \dot{\beta}}^{\ \ c}, \ \ \ \ \ T_{\alpha \dot{\beta}}^{\ \ c}=T_{\dot{\beta} \alpha}^{\ \ c}=2i \sigma_{\alpha \dot{\beta}}^{c}, \ \ \ \ \ T_{\underline{\alpha} b}^{\ \ c}=0=T_{a \underline{\beta} }^{\ \ c}, \ \ \ \ \ T_{ab}^{\ \ c}=0.
\end{split}
\end{align}
with underline denoting dotted and undotted indices. These constraints allow to solve for the connection in terms of the vielbein components,
\begin{align}\label{connection2}
\omega_{nml}=\tfrac{i}{4} e_{l a} (\psi_n \sigma^a \bar{\psi}_m-\psi_m \sigma^a \bar{\psi}_n)+\tfrac{i}{4} e_{m a} (\psi_l \sigma^a \bar{\psi}_n-\psi_n \sigma^a \bar{\psi}_l)+\tfrac{i}{4} e_{na} (\psi_l \sigma^a \bar{\psi}_m-\psi_m \sigma^a \bar{\psi}_l)& \nonumber \\
-\tfrac{1}{2} e_{l a} (\partial_n e_m^{\ a}-\partial_m e_n^{\ a})-\tfrac{1}{2} e_{ma} (\partial_l e_n^{\ a}-\partial_n e_l^{\ a})-\tfrac{1}{2} e_{na} (\partial_l e_m^{\ a}-\partial_m e_l^{\ a})&.
\end{align}

The following torsion components \cite{wessbagger}
\begin{align}\label{torsionspinor}
T_{nm}^{\ \ \alpha}|=\tfrac{1}{2} \psi_{nm}^{\ \ \alpha}, &&	T_{\gamma b}^{\ \ \alpha}|=-\tfrac{i}{6} \sigma^c_{\gamma \dot{\epsilon}} \bar{\sigma}_b^{\dot{\epsilon} \alpha} b_c-\tfrac{i}{2} \delta_\gamma^{\ \alpha} b_b, && 
T_{\ \ b}^{\dot{\gamma} \ \alpha}|=\tfrac{i}{6} \bar{\sigma}_b^{\dot{\gamma} \alpha} M.
\end{align}
together with (\ref{inverse}) allows us to compute
\begin{subequations}\label{torsions}
\begin{align}
T_{bm}^{\ \ a}|=(E_m^{\ C}T_{bC}^{\ \ a})|&=e_m^{\ c} T_{bc}^{\ \ a}+\tfrac{1}{2} \psi_m^{\ \gamma} T_{b\gamma}^{\ \ a}+\tfrac{1}{2} \bar{\psi}_{m \dot{\gamma}} T_b^{\ \dot{\gamma} a}=0, \\
T_{bm}^{\ \ \alpha}|=(E_b^{\ L} T_{Lm}^{\ \ \alpha})|&=e_b^{\ l} T_{lm}^{\ \ \alpha}|-\tfrac{1}{2} \psi_b^{\ \lambda} T_{\lambda m}^{\ \ \alpha}|-\tfrac{1}{2} \bar{\psi}_{b \dot{\lambda}} T_{\ m}^{\dot{\lambda} \ \alpha}|=e_b^{\ l} T_{lm}^{\ \ \alpha}|-\tfrac{1}{2} \psi_b^{\ \gamma} e_m^{\ c} T_{\gamma c}^{\ \ \alpha}|-\tfrac{1}{2} \bar{\psi}_{b \dot{\gamma}} e_m^{\ c} T_{\ c}^{\dot{\gamma} \ \alpha}| \nonumber \\
&=\tfrac{1}{2} e_b^{\ l} \psi_{lm}^{\ \ \alpha}|+\tfrac{i}{12} e_m^{\ c} (\psi_b^{\ \gamma} \sigma^d_{\gamma \dot{\epsilon}} \bar{\sigma}_c^{\dot{\epsilon} \alpha} b_d+3 \psi_b^{\ \alpha} b_c-\bar{\psi}_{b \dot{\gamma}} \bar{\sigma}_c^{\dot{\gamma} \alpha} M).
\end{align}
\end{subequations}

Finally, using (\ref{covder}), (\ref{doble}) and (\ref{connection2}), we have
\begin{subequations}\label{auxeq}
\begin{align}
\mathscr{D}_m \xi^a&\equiv \partial_m \xi^a+\xi^c \omega_{mc}^{\ \ \ a}=\xi^l (\partial_m e_l^{\ a}+\partial_l e_m^{\ a})+\xi^l \partial_l e_m^{\ a}+e_l^{\ a} \partial_m \xi^l+\xi^c \omega_{mc}^{\ \ \ a} \nonumber \\
&=-\tfrac{i}{2} \xi^l (\psi_l \sigma^a \bar{\psi}_m-\psi_m \sigma^a \bar{\psi}_l)-\xi^l\omega_{ml}^{\ \ a}+\xi^l\omega_{lm}^{\ \ a}+\xi^l \partial_l e_m^{\ a}+e_l^{\ a} \partial_m \xi^l+\xi^c \omega_{mc}^{\ \ \ a}, \\
\xi^l \psi_{lm}^{\ \ \alpha}&=\xi^l \partial_l \psi_m^{\ \alpha}+\psi_m^{\ \beta} \xi^l \omega_{l\beta}^{\ \ \alpha}-\partial_m (\xi^l \psi_l^{\ \alpha})+(\partial_m \xi^l) \psi_l^{\ \alpha}-\xi^l \psi_l^{\ \beta} \omega_{m \beta}^{\ \ \alpha} \nonumber \\
&=\xi^l \partial_l \psi_m^{\ \alpha}+(\partial_m \xi^l) \psi_l^{\ \alpha}-\mathscr{D}_m (\xi^l \psi_l^{\ \alpha})+\psi_m^{\ \beta} \xi^l \omega_{l\beta}^{\ \ \alpha}
\end{align}
\end{subequations}

\subsection{Wess-Zumino gauge under spacetime diffeomorphisms}\label{appb}
To simplify the following discussion, we consider a supergauge transformation\footnote{This won't affect the results of section \ref{sec3} since the Lorentz generators on the right-hand side can be redefined accordingly.}, 
\begin{align}\label{supergauge}
	\xi^A(z), && L_B^{\ A}(z) \equiv -\xi^C(z) \phi_{CB}^{\ \ \ A}(z),
\end{align}
under (\ref{supergauge}) and a further LLT, the connection transforms as \cite{wessbagger}
\begin{align}
	\delta_\xi \phi_{MB}^{\ \ \ \ A}=-\xi^C R_{CM A}^{\ \ \ \ \ B}-(-)^{m(b+c)} L_B^{\ C} \phi_{MC}^{\ \ \ A}+\phi_{MB}^{\ \ \ D} L_D^{\ A}-\partial_M L_B^{\ A}.
\end{align}
To preserve the W-Z gauge (\ref{connection}) under (\ref{diffeo}), we require that for parameters (\ref{diffeo}),
\begin{align}\label{deltaphi0}
	\delta_\xi \phi_{\mu A}^{\ \ \ B}|=(-\xi^C R_{C \mu A}^{\ \ \ \ \ B}-\partial_\mu L_B^{\ A})|=0,
\end{align}
and its complex conjugate. Terms proportional to $\xi^\alpha, \bar{\xi}_{\dot{\alpha}}$ correspond to a supergravity transformation and the required Lorentz transformation is known \cite{wessbagger}. On the other hand, the contributions from $\xi^a$ are
\begin{align}\label{deltaphi}
	0=-\xi^c R_{c\mu \alpha}^{\ \ \ \ \ \beta}|-(\partial_\mu L_\alpha^{\ \beta})|=-\tfrac{1}{2} \delta_\mu^{\ \delta} \epsilon^{\beta \gamma} \xi^c \bar{\sigma}_c^{\dot{\phi} \phi} R_{\delta \phi \dot{\phi} \alpha \gamma} -(\partial_\mu L_\alpha^{\ \beta})|,
\end{align}
where $R_{\delta \phi \dot{\phi} \alpha \gamma}$ depends on $G_{\alpha \dot{\beta}}=-\tfrac{1}{2} \sigma^c_{\alpha \dot{\beta}} G_b$, and $W_{\alpha \beta \gamma}(x)$ \cite{wessbagger}. Thus, (\ref{deltaphi0}) holds if one implements following LLT
\begin{align}
	L_{\alpha \beta}=-\tfrac{1}{2} \theta^\mu \delta_\mu^{\ \delta} \xi^c \bar{\sigma}_c^{\dot{\phi} \phi} R_{\delta \phi \dot{\phi} \alpha \beta}-\tfrac{1}{2} \bar{\theta}_{\dot{\mu}} \delta^{\dot{\mu}}_{\ \dot{\delta}} \epsilon^{\dot{\delta} \dot{\gamma}} \bar{\sigma}_c^{\dot{\phi} \phi} R_{\dot{\gamma} \phi \dot{\phi} \alpha \beta}.
\end{align}

On the other hand, the transformation law of the vielbein under (\ref{supergauge}) is
\begin{align}
	\delta_\xi E_M^{\ A}=-\mathscr{D}_M \xi^A-\xi^B T_{BM}^{\ \ \ A}.
\end{align}
Thus, to preserve the W-Z gauge (\ref{wzviel}) we require that, under the local diffeomorphism (\ref{diffeo}),
\begin{align}\label{trans}
		0&=\delta_\xi E_\mu^{\ A}|=(-\partial_\mu \xi^A-(-)^{b} \xi^B \phi_{\mu B}^{\ \ \ A}-\xi^b T_{b\mu}^{\ \ \ A}-\xi^\beta T_{\beta \mu}^{\ \ \ A}-\bar{\xi}_{\dot{\beta}} T_{\ \ \mu}^{\dot{\beta} \ A})|.
\end{align}

Evaluating (\ref{trans}) $\theta=0$ for the transformation (\ref{diffeo}), using the torsion constraints (\ref{const}), the connection (\ref{connection}), and collecting the contributions from the spinorial parameters in $\delta_\zeta$, we get
\begin{subequations}
	\begin{align}
		0&=-\partial_\mu \xi^a|+\delta_\zeta E_\mu^{\ a}|, \\
		0&=-\partial_\mu \xi^\alpha|-\xi^b \delta_\mu^{\ \gamma} T^{\ \ \alpha}_{b\gamma}|, \\
		0&=-\partial_\mu \bar{\xi}_{\dot \alpha}|-\xi^b \delta_\mu^{\ \gamma} T_{b\gamma \dot{\alpha}}|. 
	\end{align}
\end{subequations}

The contribution from the supergravity transformation (\ref{diffeo}) is canceled with $\xi^{a}_1=2i (\theta \sigma^a \bar{\zeta}-\zeta \sigma^a \bar{\theta})$ \cite{wessbagger}. To get rid of the contributions from $\xi^a$, we also require
\begin{align}
\xi_1^\alpha=-\xi^b \theta^\mu \delta_\mu^{\ \gamma} T^{\ \ \alpha}_{b\gamma}|-\xi^b \bar{\theta}_{\dot \mu} \delta^{\dot \mu}_{\ \dot{\gamma}} T_b^{\ \dot{\gamma} \alpha}|,
\end{align}
with the torsion components given in (\ref{torsionspinor}).

\subsection{Auxiliary fields}\label{appc}
They are the lowest components of a chiral superfield $R(z)$ ($\bar{\mathscr{D}}^{\dot{\alpha}} R=0$) and a vector superfield $G_a(z)$ (related to $R$ by $\mathscr{D}_\alpha R=\mathscr{D}^{\dot{\alpha}} G_{\alpha \dot{\alpha}}$) parametrizing the solutions of the Bianchi identities \cite{wessbagger},
\begin{align}\label{auxiliary}
M(x) \equiv -6 R(z)|, && b_a(x)\equiv -3 G_a(z)|.
\end{align}

For the scalar field, (\ref{auxiliary}), we have  
\begin{align}\label{deltar}
\delta_\xi M&=-\xi^c \mathscr{D}_c (-6 R(z))|=6 \xi^c (E_c^{\ L} \mathscr{D}_L R)| \nonumber \\
&=6 \xi^c (e_c^{\ l} \mathscr{D}_l R|-\tfrac{1}{2} \psi_c^{\ \lambda} \delta_\lambda^{\ \alpha} \mathscr{D}_\alpha R|)=-\xi^l \partial_l M-\tfrac{3}{2} \xi^l \psi_l^{\ \alpha} \mathscr{D}_\alpha R|.
\end{align}
Using the value of the covariant derivative $\mathscr{D}_\alpha R|$ given in \cite{wessbagger}, yields (\ref{newscal}).

On the other hand, for a vector superfield, we have, using (\ref{covder}),
\begin{align}
\delta_\xi G_a\equiv -\xi^L \partial_L G_a=-\xi^L \mathscr{D}_L G_a-\xi^L \phi_{La}^{\ \ b} G_b=-\xi^C \mathscr{D}_C G_a+G_b \xi^L \phi_{L\ a}^{\ b}=-\xi^c \mathscr{D}_c G_a+G_b \xi^L \phi_{L\ a}^{\ b}.
\end{align}
Then,
\begin{align}\label{vectoraux}
\delta_\xi b_a&=3 \xi^c \mathscr{D}_c G_a|+b_b \xi^l \omega_{l\ a}^{\ b}=3 \xi^c (E_c^{\ L} \mathscr{D}_L G_a)|+b_b \xi^l \omega_{l\ a}^{\ b} \nonumber \\
&=3 \xi^c (e_c^{\ l} \mathscr{D}_l G_a|-\tfrac{1}{2} \psi_c^{\ \lambda} \delta_{\lambda}^{\ \delta} \mathscr{D}_\delta G_a-\tfrac{1}{2} \bar{\psi}_{c\dot{\lambda}} \delta^{\dot{\lambda}}_{\ \ \dot{\delta}} \bar{\mathscr{D}}^{\dot{\delta}} G_a|)+b_b \xi^l \omega_{l\ a}^{\ b}.
\end{align}	
Expanding the covariant derivative we get (\ref{newvector}).

\ \ \\

\centerline{\bf Acknowledgements}
N.E. Martínez-Pérez thanks CONAHCyT for financial support.

\bibliography{isometries}

\end{document}